\documentclass[final,5p,times,twocolumn,authoryear]{elsarticle}

\usepackage{amssymb}
\usepackage{lipsum}

\newcommand{\mnras}{Mon. Not. R. Astron. Soc.}
\newcommand{\apjl}{Astrophys. J. Lett.}
\newcommand{\apj}{Astrophys. J.}
\newcommand{\apjs}{Astrophys. J. Suppl. Ser.}
\newcommand{\aap}{Astron. \& Astrophys.}
\newcommand{\physrep}{Physics Reports}
\newcommand{\nar}{New Astronomy Reviews}
\newcommand{\prl}{Phys. Rev. Lett.}

\journal{High Energy Astrophysics}

\begin{document}

\begin{frontmatter}



\title{\bf The BOAT GRB 221009A: a Poynting-Flux-Dominated Narrow Jet Surrounded by a Matter-Dominated Structured Jet Wing}


\author[1,2]{Bing Zhang}
\ead{bing.zhang@unlv.edu}
\affiliation[1]{organization={Nevada Center for Astrophysics, University of Nevada Las Vegas},
            city={Las Vega },
            postcode={89154}, 
            state={NV},
            country={USA}}
\affiliation[2]{organization={Department of Physics and Astronomy, University of Nevada Las Vegas},
            city={Las Vega },
            postcode={89154}, 
            state={NV},
            country={USA}}
\author[3,1]{Xiang-Yu Wang}
\ead{xywang@nju.edu.cn}
\author[3]{Jian-He Zheng}
\affiliation[3]{organization={School of Astronomy and Space Science, Nanjing University},
            city={Nanjing},
            postcode={210093}, 
            country={China}}

\begin{abstract}
We argue that the broad-band observations of the brightest-of-all-time (BOAT) GRB 221009A reveal a physical picture involving two jet components: a narrow ($\sim 0.6$ degree half opening angle) pencil-beam jet that has a Poynting-flux-dominated jet composition, and a broader matter-dominated jet with an angular structure. We discuss various observational evidence that supports such a picture. To treat the problem, we develop an analytical structured jet model for both forward and reverse shock emission from the matter dominated structured jet wing during the deceleration phase. We discuss the physical implications of such a two-component jet configuration for this particular burst and for GRBs in general. We argue that in a quasi-universal structured jet scenario, some bright X-ray flares could be similar narrow jets viewed slightly outside the narrow jet cone and that narrow jets may exist in many more GRBs without being detected.
\end{abstract}




\end{frontmatter}




\section{Introduction}
\label{sec:intro}

GRB 221009A is the ``brightest-of-all-time'' (BOAT) \citep{BOAT-burns}. Its fluence and peak flux are the highest ever detected by the humanity since the development of gamma-ray astronomy. At redshift $z=0.151$ \citep{BOAT-z=0.151}, its isotropic energy $E_{\rm iso} \sim 1.5\times 10^{55}$ erg \citep{BOAT-an,BOAT-Konus} is also the largest ever detected and its peak isotropic luminosity is at the $\sim$ 99th percentile of the known distribution. Despite its enormously large apparent energetics, its prompt emission and afterglow properties are consistent with the known long GRB distributions extending to the high-energy end \citep{BOAT-lan,BOAT-kann}, suggesting that it does not demand a brand new type of progenitor system different from that of known long GRBs.

One unique aspect of this burst was its detection in the TeV range (up to above 10 TeV) during and shortly after the prompt emission phase of the MeV emission by the LHAASO telescope \citep{BOAT-cao,BOAT-cao2}. Two unique lightcurve features were detected. First, a clear lightcurve peak was detected around $\sim T_* + 18$ s after the zero time $T_* = T_0 + 226$ s of the afterglow onset ($T_0$ is the GBM trigger time), which can be interpreted as the deceleration time of the external shock. This break is detected for the first time in the synchrotron self-Compton (SSC) regime. For a constant density medium, this gives a measurement of the bulk Lorentz factor of the jet as\footnote{We adopt Eq.(7.81) of \cite{zhang18} which includes a more precise integration and is about a factor of 1.31 times of Eq.(1) of \cite{molinari07}. }
\begin{eqnarray}
    \Gamma_0 & = & 0.9^{3/8} \left( \frac{3E(1+z)^3}{2\pi \hat\gamma n m_p c^5 t_{\rm dec}^3}\right)^{1/8} \nonumber \\ 
    &\simeq & 600 \left(\frac{t_{\rm dec}}{18 \ {\rm s}}\right)^{-3/8} \left(\frac{1+z}{1.151}\right)^{3/8} E_{55}^{1/8}n^{-1/8}. 
\label{eq:Gamma0}
\end{eqnarray}
Second, a clear ``jet break'' was detected in the TeV band at $\sim T_*+670$ s after the onset of the afterglow \citep{BOAT-cao}. This defines a half jet opening angle of
\begin{eqnarray}
    \theta_{\rm j} & \simeq & (0.01 \ {\rm rad} ~{\rm or}~ 0.6^\circ) \left(\frac{t_{\rm j}}{670 \ {\rm s}}\right)^{3/8} \left(\frac{1+z}{1.151}\right)^{-3/8} \nonumber \\ 
    &\times& \left(\frac{E_{\rm \gamma,iso}}{1.5\times 10^{55} \ {\rm erg}}\right)^{-1/8} (\eta_\gamma n)^{1/8}
\label{eq:thetaj}
\end{eqnarray}
for a constant density medium, where $\eta_\gamma$ is the ratio between $\gamma$-ray and kinetic energies, and $n$ is the ambient number density in units of $\rm cm^{-3}$, which are left as free parameters. Such a small jet opening angle brings the true emitted $\gamma$-ray energy down to $7.5\times 10^{50} \ {\rm erg}$, which falls right in the middle of the $E_\gamma$ distribution of long GRBs \citep{frail01,wangxg18}. The combined GECAM and Insight-HXMT MeV lightcurve also revealed an early jet break that is consistent with such a picture \citep{BOAT-an}. 

The lower-energy, later-time broadband afterglow, on the other hand, suggests a different picture. The simplest constant energy, constant density forward shock afterglow model was found to have difficulty to account for the broad-band afterglow emission \citep[e.g.][]{BOAT-laskar,BOAT-kann} and a structured jet model, which has long been introduced to interpret GRB properties \citep{meszaros98,daigou01,zhangmeszaros02b,rossi02,kumargranot03,granotkumar03} and was proven to be essential to interpret the gravitational-wave-associated short GRB 170817A \citep[e.g.][]{lazzati18,troja20,ioka21,beniamini20}, was found necessary to account for the broad-band data \citep{BOAT-structured-jet,BOAT-Gill}. A two-component jet model invoking a narrow top-hat jet and a broader top-hat jet was speculated by \cite{BOAT-sato}. However, the model parameters derived by various groups to fit the late-time afterglow data \citep[e.g.][]{BOAT-ren-wind,BOAT-laskar,BOAT-structured-jet,BOAT-Gill,BOAT-sato,BOAT-zhang-proton} are found unable to simultaneously account for the early LHAASO data \citep{zheng23}. 

The purpose of this Letter is not to provide a global fit to the broad-band (including TeV emission) afterglow data of GRB 221009A. Such a task is being carried out by several groups independently \citep{zheng23,ren23,wangk23,BOAT-zhang2}. Rather, this Letter is meant to provide an outline of a novel physical picture that may account for various observations of this burst, which may be helpful to guide the development of more detailed afterglow models to fit the data. A general two-component jet theoretical framework is presented in Section \ref{sec:framework}. An analytical on-axis structured jet model is developed in Section \ref{sec:structured-jet}. The arguments in favor of a narrow Poynting-flux-dominated pencil beam jet and a wider matter-dominated structured jet wing are discussed in Section \ref{sec:2-comp}. Broader implications of such a two-component jet model are discussed in Section \ref{sec:implications}, with the suggestion that some giant X-ray flares could be narrow jet emission viewed slightly outside the jet cone. The results are summarized in Section \ref{sec:summary}. 

\section{Theoretical framework}\label{sec:framework}
\begin{figure*}[t]
\centering
\includegraphics[width=1.65\columnwidth]{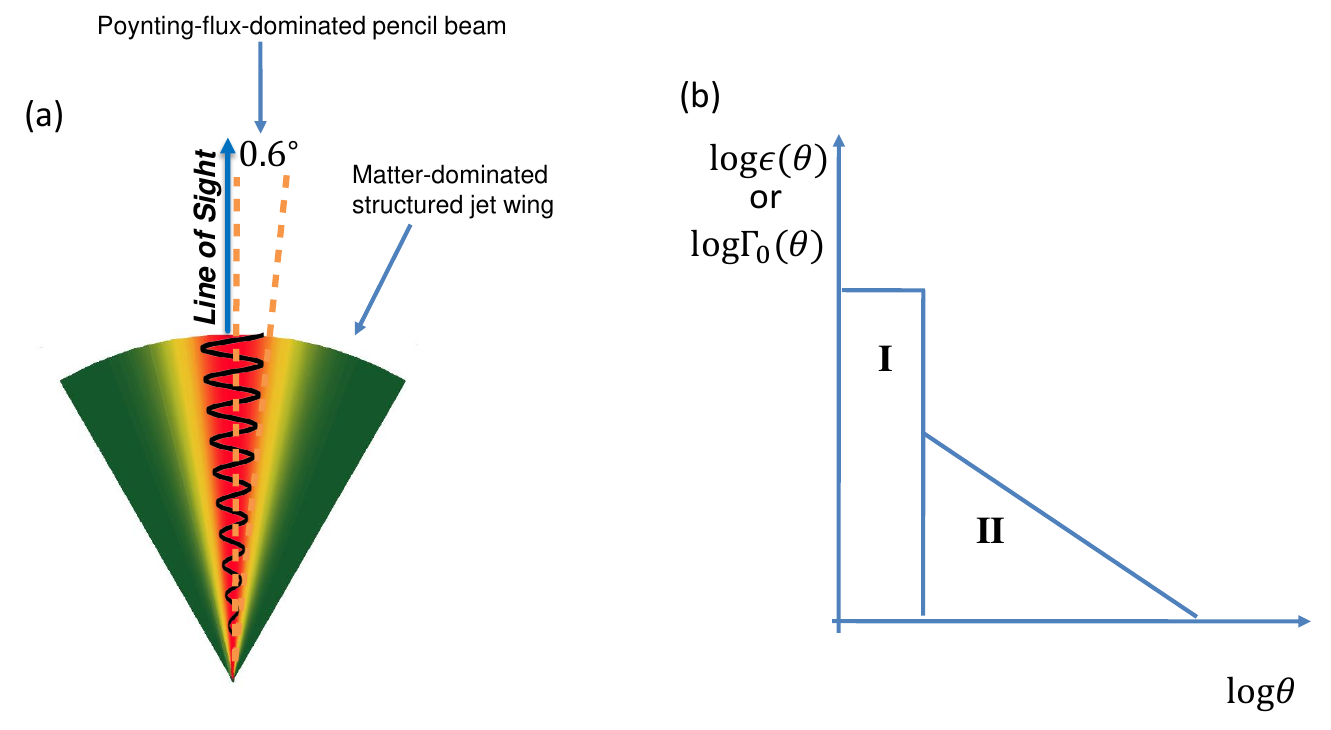}
\caption{(a) A cartoon picture of the proposed two-component jet model. The central pencil beam has roughly a uniform profile with sharp edge, which is Poynting flux dominated. The wing has an angular structured and is matter dominated. (b) An illustration of the energy per solid angle and Lorentz factor profile of the proposed jet.}
\label{fig:cartoon}
\end{figure*}

We propose the following two-component jet model to interpret GRB 221009A (see  Figure \ref{fig:cartoon} for a cartoon picture). The jet includes a narrow pencil beam with an opening angle of $\theta_{\rm j} \simeq 0.6^\circ$ that is Poynting flux dominated (denoted as Jet I) and an outside structured jet wing that is matter dominated (denoted as Jet II). The two components are distinct in energy distribution, i.e. there is a step-function jump between the isotropic energy per solid angle at the edge of the narrow jet with angle $\theta_{\rm j}$. This is required to interpret the early jet break in the LHAASO data \citep{BOAT-cao}. Mathematically, the angle-dependent energy per solid angle is defined as

\begin{eqnarray}
    \epsilon \equiv \frac{dE}{d\Omega} = \left \{
      \begin{array}{ll}
        \epsilon_{\rm I}, & \theta < \theta_{\rm j}, \\
        \epsilon_{\rm II} f(\theta), & \theta_{\rm j} < \theta < \Theta,
      \end{array}
      \right.
\label{eq:epsilon-structure}
\end{eqnarray}
where $\theta_{\rm j}$ is the opening angle of the narrow jet, $\Theta$ is the maximum angle of the structured jet, $\epsilon_{\rm I} \gg \epsilon_{\rm II}$ and $f(\theta)$ is the function form (e.g. a power law, a Gaussian, or any other arbitrary structure, \citealt{zhangmeszaros02b,rossi02}) of the structured jet. Correspondingly, one could also define a jet structure for an angle-dependent initial Lorentz factor, i.e.
\begin{eqnarray}
    \Gamma_0(\theta) = \left \{
      \begin{array}{ll}
        \Gamma_{\rm I}, & \theta < \theta_{\rm j} \\
        \Gamma_{\rm II} g(\theta), & \theta_{\rm j} < \theta < \Theta,
      \end{array} 
      \right.
\label{eq:Gamma-structure}
\end{eqnarray}    
where $\Gamma_{\rm I} \gg \Gamma_{\rm II}$ and $g(\theta)$ is the structure function of the Lorentz factor profile, which generally is different from $f(\theta)$. As an example, the power law structured jet in the  $ \bf {\theta \gg \theta_{\rm j}}$ regime can be characterized by
\begin{eqnarray}
   f(\theta) & = & \left(\frac{\theta}{\theta_{\rm j}}\right)^{-k_{\epsilon}}, \\
   g(\theta) & = & \left(\frac{\theta}{\theta_{\rm j}}\right)^{-k_{\Gamma}}.
\end{eqnarray}
We will discuss in \S\ref{sec:2-comp} the detailed arguments in favor of such a jet structure to interpret this GRB. In the following, we first develop an analytical model of an on-axis structured jet with both forward shock (FS) and reverse shock (RS) emission.

\section{An analytical model for on-axis structured jets}\label{sec:structured-jet}

\subsection{Forward shock emission}\label{sec:FS}

One important parameter in the standard afterglow modeling is the isotropic-equivalent energy $E_{\rm iso}$. For a uniform (top-hat) jet with $\epsilon=$ const for $\theta<\theta_{\rm j}$ and for a on-beam observer, one has $E_{\rm iso} = 4\pi \epsilon$. For a structured jet, the situation is more complicated. For an off-axis observer with viewing angle $\theta_v$, one has $E_{\rm iso} \sim 4\pi \epsilon(\theta_v)$ when $[\Gamma(\theta_v)]^{-1} \ll \theta_v$ \citep{zhangmeszaros02b}, but the case is more complicated and depends on the details of the structure when $[\Gamma(\theta_v)]^{-1} \sim \theta_v$ \citep{kumargranot03,granotkumar03}. A precise description requires numerical calculations. 
    
For an on-axis observer like the case of GRB 221009A, an analytical treatment is possible. For the jet structure specified in Eqs. (\ref{eq:epsilon-structure}) and (\ref{eq:Gamma-structure}), one can define the effective isotropic energies within an angle $\theta$ from the line of sight (which is also the jet axis) of the two jet components, i.e. 
\begin{eqnarray}
    E_{\rm iso}^{\rm I} (\theta)
    &=& \left\{
     \begin{array}{ll}
        4\pi \epsilon_{\rm I}, & \theta < \theta_{\rm j} \\
        0,  & \theta_{\rm j}<\theta<\Theta.
     \end{array}  \label{eq:EisoI} 
    \right.
\end{eqnarray}

\begin{eqnarray}
    \bar E_{\rm iso}^{\rm II} (\theta)
     &=& \left\{
     \begin{array}{ll}
        0, & \theta < \theta_{\rm j}, \\
        \frac{4\pi \int_{\theta_{\rm j}}^{\theta} \epsilon_{\rm II} f(\theta) \sin\theta d \theta}{1-\cos\theta}, & \theta_{\rm j}<\theta<\Theta,
     \end{array}  
     \right. \nonumber \\
     & \simeq & \left\{
     \begin{array}{ll}
        0, & \theta < \theta_{\rm j}, \\
        \frac{8\pi \epsilon_{\rm II}}{2-k_\epsilon} \left[ \left(\frac{\theta_{\rm j}}{\theta}\right)^{k_\epsilon}  -\left(\frac{\theta_{\rm j}}{\theta}\right)^2 \right], & \theta_{\rm j}<\theta<\Theta. (k_\epsilon \neq 2), \\
        8\pi \epsilon_{\rm II} \ln \left(\frac{\theta}{\theta_{\rm j}}\right)\left(\frac{\theta_{\rm j}}{\theta}\right)^2, & \theta_{\rm j}<\theta<\Theta. (k_\epsilon = 2). 
     \end{array}
    \right.  \nonumber \\
\label{eq:Eiso2}
\end{eqnarray}
Note that in this treatment, the second component only has the wing component beyond $\theta_{\rm j}$ and it is normalized within the entire $\theta$-cone. We consider an on-axis observer. The effective isotropic energy is calculated by the total energy in the solid angle between $\theta_{\rm j}$ and $\theta = 1/\Gamma$ normalized to the solid angle from 0 to $\theta$ and multiplied by $4\pi$. As a result, the contribution from the second component is zero before the wing jet rings are progressively decelerated to enter the $1/\Gamma$ cone of the observer. Also note that in the second approximation of (Eq. \ref{eq:Eiso2}) the power law structured jet model with $\theta \ll 1$ has been adopted.

It is worth noticing that $E_{\rm iso}^{\rm II}$ is not a monotonic function of $\theta$. It starts with 0 at $\theta=\theta_{\rm j}$, reaches a maximum at 
\begin{equation}
    \theta_c = \left\{
    \begin{array}{ll}
      \left(\frac{2}{k_\epsilon}\right)^{\frac{1}{2-k_\epsilon}}\theta_{\rm j},  & k_\epsilon \neq 2 \\
      \sqrt{e} \theta_{\rm j},   & k_\epsilon = 2.
    \end{array} \right.
\end{equation}
and then drops as a power law at $\theta \gg \theta_c$, with
\begin{equation}
 \bar E_{\rm iso}^{\rm II} (\theta) \propto \theta^{-a} \end{equation}
where
\begin{equation}
 a = \left\{
 \begin{array}{ll}
   k_\epsilon, & k_{\epsilon} \leq 2, \\
   2, & k_{\epsilon} > 2.
 \end{array}
 \right.
\end{equation}
This suggests that the emission from Jet II is initially not very strong and only becomes significant when $\theta$ reaches $\theta_c$. 
If one includes both jet components, the effective energy per solid angle $E_{\rm iso}(\theta)$ decreases monotonically with $\theta$ once $\theta > \theta_{\rm j}$. This is the opposite case from the energy injection model. It tends to make the decay index steeper than the isotropic fireball case but is shallower than the post-jet-break case.
 
\subsubsection{Constant density medium}

For a jet with an isotropic energy $E_{\rm iso}$ running into a constant density medium $n$, the Lorentz factor evolution during the self-similar deceleration phase evolves as \citep{meszarosrees97,sari98}
\begin{equation}
    \Gamma(t) \propto E_{\rm iso}^{1/8} n^{-1/8} t^{-3/8}
\label{eq:Gamma(t)}
\end{equation}
As the jet decelerates from the core region, one observes a progressively larger solid angle defined by $\theta \sim \Gamma^{-1}$ as $\Gamma$ continues to decrease. If the power-law Lorentz factor profile $g(\theta)$ has $k_{\Gamma} \leq 1$, the annular ring at $\theta$ has already decelerated when it enters the field of view, so the blastwave dynamics is independent of $k_\Gamma$. This is not the case for $k_\Gamma> 1$. In the following, we consider the two cases separately.

\begin{center}
{Case I: $k_\Gamma\le 1$}
\end{center}

Let $\Gamma = \theta^{-1}$ and use the function $\bar E^{\rm II}_{\rm iso} (\theta)$ and Eq.(\ref{eq:Gamma(t)}), one can define the dynamics of the blastwave for the case of $k_\Gamma \leq 1$. Let us adopt the approximation 
\begin{equation}
 E_{\rm iso} (\theta) \simeq \bar E_{\rm iso}^{\rm II} (\theta) \propto \theta^{-a} \propto \Gamma^a
\end{equation}
for the asymptotic values at $\theta \gg \theta_{\rm j}$,
one can derive $\Gamma(t) \propto t^{-3/(8-a)}$,  
$E_{\rm iso}(t) \propto t^{-3a/(8-a)}$,  $r \propto t^{(2-a)/(8-a)}$, $F_{\rm \nu.max} \propto t^{-3a/(8-a)}$, $\nu_m \propto \Gamma^3 B' \propto t^{-12/(8-a)}$ ($B' \propto \Gamma$), and $\nu_c \propto \Gamma^{-1}t^{-2} {B'}^{-3} \propto t^{(2a-4)/(8-a)}$. For the self-absorption frequency, we have $\nu_a \propto E_{\rm iso}^{1/5} \propto t^{-3a/5(8-a)}$ for $\nu_a < \nu_m < \nu_c$, $\nu_a \propto E_{\rm iso}^{7/10} t^{-1/2} \propto t^{-(8a+20)/5(8-a)}$ for $\nu_a < \nu_c < \nu_m$, and $\nu_a \propto E_{\rm iso}^{(p+2)/2(p+4)} t^{-(3p+2)/2(p+4)} \propto t^{-(12p+2a+8)/(p+4)(8-a)}$ for $\nu_m < \nu_a < \nu_c$ (we limit ourselves only to the weak-absorption case with $\nu_a < \nu_c$).
This finally gives 
\begin{equation}
F_\nu \propto \left\{
 \begin{array}{ll}
   t^{\frac{4-2a}{8-a}}, & \nu < \nu_a \\
   t^{\frac{4-3a}{8-a}}, & \nu_a < \nu < \nu_m \\
   t^{\frac{-6p-3a+6}{8-a}}, & \nu_m < \nu < \nu_c \\
   t^{\frac{-6p-2a+4}{8-a}}, & \nu > \nu_c 
 \end{array}
    \right.
\end{equation}
or
\begin{equation}
F_\nu \propto \left\{
 \begin{array}{ll}
   t^{\frac{4-2a}{8-a}}, & \nu < \nu_m \\
   t^{\frac{10-2a}{8-a}}, & \nu_m < \nu < \nu_a \\
   t^{\frac{-6p-3a+6}{8-a}}, & \nu_m < \nu < \nu_c \\
   t^{\frac{-6p-2a+4}{8-a}}, & \nu > \nu_c 
 \end{array}
    \right.
\end{equation}
for slow cooling ($\nu_m < \nu_c$) and 
\begin{equation}
F_\nu \propto \left\{
 \begin{array}{ll}
   t^{1}, & \nu < \nu_a  \\
   t^{\frac{4-11a}{3(8-a)}}, & \nu_a < \nu < \nu_c \\
   t^{\frac{-2(a+1)}{8-a}}, & \nu_c < \nu < \nu_m \\
   t^{\frac{-6p-2a+4}{8-a}}, & \nu > \nu_m 
 \end{array}
    \right.
\end{equation}
for fast cooling ($\nu_c < \nu_m$). Note that these scaling laws are generally in the same form as previous treatments \citep{meszaros98,panaitescu05,beniamini22,BOAT-structured-jet}. However, in contrast to previous structured jet treatments that adopt $E_{\rm iso}(\theta) = 4\pi \epsilon(\theta)$, we adopted the effective isotropic energy $\bar E_{\rm iso}^{\rm II} (\theta)$ defined in Eq.(\ref{eq:Eiso2}) that includes {\em integrated energy normalized to the solid angle within angle $\theta$}. It turns out that for power-law jet models with $k_\epsilon \leq 2$, at large $\theta$ angles, $\bar E_{\rm iso}(\theta)$ and $\epsilon(\theta)$ have the same power law index. At a smaller angle with $\theta \gtrsim \theta_{\rm j}$, the simple power law treatment does not apply. Our treatment avoids the unreasonably large $E_{\rm iso}$ value when the second jet component just enters the line of sight and allows a more accurate treatment. Also for the case of $k_\epsilon > 2$, our treatment gives $a=2$ rather than $a=k_\epsilon$, which provides a more accurate description of the blastwave dynamics.

\begin{center}
{Case II: $k_\Gamma> 1$}    
\end{center}

For a Lorentz factor profile $k_\Gamma > 1$, one needs to plug  Eq.(\ref{eq:Eiso2}) into Eq.(\ref{eq:Gamma(t)}) and let it equal to Eq.(\ref{eq:Gamma-structure}) and consequently solve for $\Gamma(t)$ and $E_{\rm iso}(t)$ self-consistently. 
In this case, we have $\Gamma(t)=\Gamma_0(\theta)\propto \theta^{-k_{\Gamma}}$ and $\theta\propto \Gamma^{-1/k_\Gamma}$, so that
\begin{equation}
    E_{\rm iso}(\theta) \simeq \bar E_{\rm iso}^{\rm II}\propto\theta^{-a}\propto \Gamma^{a/k_{\Gamma}}
\label{eq:E_ism(k>1)}
\end{equation}
Substituting Eq.(\ref{eq:E_ism(k>1)}) into Eq.(\ref{eq:Gamma(t)}), one  obtains 
\begin{equation}
\Gamma(t)\propto t^{\frac{-3k_\Gamma}{8k_\Gamma-a}},
\end{equation}
which gives
    \begin{eqnarray}
         E_{\rm iso}(t) & \propto & t^{-\frac{3a}{8 k_\Gamma-a}}, \\
        \theta & \propto &t^{\frac{3}{8 k_\Gamma-a}}, \\
        r(t) & \propto & t^{\frac{2k_\Gamma-a}{8k_\Gamma-a}}.
    \end{eqnarray}
From these scaling laws, one finds that in the $k_\Gamma > 1$ case, all the scaling laws are the same as the case of $k_\Gamma \leq 1$ if we replace $a$ with a new variable 
\begin{equation}
    A \equiv a / k_\Gamma.
\end{equation} 
This  gives
\begin{equation}
F_\nu \propto \left\{
 \begin{array}{ll}
   t^{\frac{4-2A}{8-A}}, & \nu < \nu_a \\
   t^{\frac{4-3A}{8-A}}, & \nu_a < \nu < \nu_m \\
   t^{\frac{-6p-3A+6}{8-A}}, & \nu_m < \nu < \nu_c \\
   t^{\frac{-6p-2A+4}{8-A}}, & \nu > \nu_c 
 \end{array}
    \right.
\end{equation}
or
\begin{equation}
F_\nu \propto \left\{
 \begin{array}{ll}
   t^{\frac{4-2A}{8-A}}, & \nu < \nu_m \\
   t^{\frac{10-2A}{8-A}}, & \nu_m < \nu < \nu_a \\
   t^{\frac{-6p-3A+6}{8-A}}, & \nu_m < \nu < \nu_c \\
   t^{\frac{-6p-2A+4}{8-A}}, & \nu > \nu_c 
 \end{array}
    \right.
\end{equation}
for slow cooling ($\nu_m < \nu_c$) and 
\begin{equation}
F_\nu \propto \left\{
 \begin{array}{ll}
   t^{1}, & \nu < \nu_a \\
   t^{\frac{4-11A}{3(8-A)}}, & \nu_a < \nu < \nu_c \\
   t^{\frac{-2(A+1)}{8-A}}, & \nu_c < \nu < \nu_m \\
   t^{\frac{-6p-2A+4}{8-A}}, & \nu > \nu_m 
 \end{array}
    \right.
\end{equation}
for fast cooling ($\nu_c < \nu_m$). All the scaling goes back to the standard case \citep{Gaohe2013} when $a=0$.

\subsubsection{Wind medium}
For a pre-explosion stellar wind with a constant velocity, the density can be characterized as $n=A_w r^{-2}$ \citep{dailu98c,chevalier00}, where $A_w =\dot{M}/(4\pi v_w m_p)$, $\dot{M}$ is the mass-loss rate of the massive star that ejected the wind at a constant speed $v$. In a wind medium, the Lorentz factor evolution in the self-similar deceleration phase reads
\begin{equation}
    \Gamma(t) \propto E_{\rm iso}^{1/4} A_w^{-1/4} t^{-1/4}.
\label{eq:Gamma(t)-wind}
\end{equation}
Again we discuss two cases.

\begin{center}
{Case I: $k_\Gamma\le 1$}
\end{center}

Let $\Gamma = \theta^{-1}$ and use the relation $E_{\rm iso} (\theta) \propto \theta^{-a} \propto \Gamma^a$ and Eq.(\ref{eq:Gamma(t)-wind}), one can derive
$\Gamma(t) \propto t^{-1/(4-a)}$,  
$E_{\rm iso}(t) \propto t^{-a/(4-a)}$,  $r \propto t^{(2-a)/(4-a)}$, $F_{\rm \nu.max} \propto t^{-2/(4-a)}$, $\nu_m \propto t^{-(6-a)/(4-a)}$, and $\nu_c \propto t^{(2-a)/(4-a)}$ (noticing $B' \propto \Gamma r^{-1}$). For synchrotron self-absorption, we have $\nu_a \propto E_{\rm iso}^{-2/5}t^{-3/5} \propto t^{(5a-12)/5(4-a)}$ for $\nu_a < \nu < \nu_m$, $\nu_a \propto E_{\rm iso}^{-2/5} t^{-8/5} \propto t^{(10a-32)/5(4-a)}$ for $\nu_a < \nu_c < \nu_m$, and $\nu_a \propto E_{\rm iso}^{(p-2)/2(p+4)} t^{-3(p+2)/2(p+4)}\propto t^{-(6p-4a-ap+12)/(p+4)(4-a)}$ for $\nu_m < \nu_a < \nu_c$.
This gives 
\begin{equation}
F_\nu \propto \left\{
 \begin{array}{ll}
   t^{\frac{4-2a}{4-a}}, & \nu < \nu_a \\
   t^{-\frac{a}{3(4-a)}}, & \nu_a < \nu < \nu_m \\
   t^{-\frac{(6-a)p+a-2}{2(4-a)}}, & \nu_m < \nu < \nu_c \\
   t^{-\frac{(6-a)p+2a-4}{2(4-a)}}, & \nu > \nu_c 
 \end{array}
    \right.
\end{equation}
or
\begin{equation}
F_\nu \propto \left\{
 \begin{array}{ll}
   t^{\frac{4-2a}{4-a}}, & \nu < \nu_m \\
   t^{\frac{14-5a}{2(4-a)}}, & \nu_m < \nu < \nu_a \\
   t^{-\frac{(6-a)p+a-2}{2(4-a)}}, & \nu_a < \nu < \nu_c \\
   t^{-\frac{(6-a)p+2a-4}{2(4-a)}}, & \nu > \nu_c 
 \end{array}
    \right.
\end{equation}
for slow cooling ($\nu_m < \nu_c$) and 
\begin{equation}
F_\nu \propto \left\{
 \begin{array}{ll}
   t^{\frac{8-3a}{4-a}}, & \nu < \nu_a \\
   t^{-\frac{8-a}{3(4-a)}}, & \nu_a < \nu < \nu_c \\
   t^{-\frac{2+a}{2(4-a)}}, & \nu_c < \nu < \nu_m \\
   t^{-\frac{(6-a)p+2a-4}{2(4-a)}}, & \nu > \nu_m 
 \end{array}
    \right.
\end{equation}
for fast cooling ($\nu_c < \nu_m$).

\begin{center}
{Case II: $k_\Gamma> 1$}    
\end{center}

Same as the constant density case, for a Lorentz factor profile $k_\Gamma > 1$, one has $\Gamma\propto \theta^{-k_{\Gamma}}$, $\theta\propto \Gamma^{-1/k_\Gamma}$, and 
\begin{equation}
    E_{\rm iso}(\theta)\propto\theta^{-a}\propto \Gamma^{a/k_{\Gamma}}.
\label{eq:E_iso(k>1)}
\end{equation} 
All the scaling relations in the $k_\Gamma \leq 1$ case can be applied by replacing $a$ by $A=a/k_\Gamma$. This gives 
$E_{\rm iso} \propto t^{-A/(4-A)}$, $r \propto t^{(2-A)/(4-A)}$, and
\begin{equation}
F_\nu \propto \left\{
 \begin{array}{ll}
   t^{\frac{4-2A}{4-A}}, & \nu < \nu_a \\
   t^{-\frac{A}{3(4-A)}}, & \nu_a < \nu < \nu_m \\
   t^{-\frac{(6-A)p+A-2}{2(4-A)}}, & \nu_m < \nu < \nu_c \\
   t^{-\frac{(6-A)p+2A-4}{2(4-A)}}, & \nu > \nu_c 
 \end{array}
    \right.
\end{equation}
or
\begin{equation}
F_\nu \propto \left\{
 \begin{array}{ll}
   t^{\frac{4-2A}{4-A}}, & \nu < \nu_m \\
   t^{\frac{14-5A}{2(4-A)}}, & \nu_m < \nu < \nu_a \\
   t^{-\frac{(6-A)p+A-2}{2(4-A)}}, & \nu_a < \nu < \nu_c \\
   t^{-\frac{(6-A)p+2A-4}{2(4-A)}}, & \nu > \nu_c 
 \end{array}
    \right.
\end{equation}
for slow cooling ($\nu_m < \nu_c$) and 
\begin{equation}
F_\nu \propto \left\{
 \begin{array}{ll}
   t^{\frac{8-3A}{4-A}}, & \nu < \nu_a \\
   t^{-\frac{8-A}{3(4-A)}}, & \nu_a < \nu < \nu_c \\
   t^{-\frac{2+A}{2(4-A)}}, & \nu_c < \nu < \nu_m \\
   t^{-\frac{(6-A)p+2A-4}{2(4-A)}}, & \nu > \nu_m 
 \end{array}
    \right.
\end{equation}
for fast cooling ($\nu_c < \nu_m$). Again all the scaling reduces to the standard case when $A=0$ \citep{Gaohe2013}.

We summarize all the spectral and temporal indices (closure relations) for both FS and RS emission in Table \ref{tab:relation}.

\begin{table*}[t]
\centering
\caption{Temporal index $\alpha$ and spectral index
$\beta$ for the forward shock and reverse shock emission in the structured jet model. The convention $F_\nu \propto t^{\alpha}\nu^{\beta}$ has been adopted. \label{Table1}} 

\begin{tabular}{llllll}
\hline\hline  
& &\hskip 1cm {\rm  forward shock}  &  & \hskip 1cm {\rm reverse shock }  & \\

& $\beta$ & $\alpha (k_\Gamma \leq 1) $  &  $\alpha (k_\Gamma>1)$ & $\alpha (k_\Gamma \leq 1)$ & $\alpha (k_\Gamma>1)$  \\
\hline
ISM & slow cooling \\
\hline
$\nu<\nu_a$   &  ${2}$  &   $\frac{4-2a}{8-a}$ & $\frac{4-2A}{8-A}$ & $ -\frac{2(14a-22k_\Gamma-27)}{7(15+k_\Gamma-2a)}$ & $ \frac{7-2A}{8-A}$\\
$\nu_a<\nu<\nu_m$   &  ${1 \over 3}$  &   $\frac{4-3a}{8-a}$ & $\frac{4-3A}{8-A}$ & $-\frac{2(24-17k_\Gamma+21a)}{7(15+k_\Gamma-2a)}$ & $-\frac{3A+1}{8-A}$\\
$\nu_m<\nu<\nu_c$   &  ${-{p-1 \over 2}}$   &  $\frac{-6p-3a+6}{8-a}$
  & $\frac{-6p-3A+6}{8-A}$ & $-\frac{3(7-7k_\Gamma+14a+27p-13k_\Gamma p)}{7(15+k_\Gamma-2a)}$ & $-\frac{3(A+p)}{8-A}$\\
$\nu>\nu_c$   &  ${-{p\over 2}}$   &   $\frac{-6p-2a+4}{8-a}$  & $\frac{-6p-2A+4}{8-A}$ & ${\rm cutoff}$ & {\rm cutoff} \\
\hline
$\nu<\nu_m$   &  ${2}$  &   $\frac{4-2a}{8-a}$ & $\frac{4-2A}{8-A}$ &  $\frac{2(27+22 k_\Gamma-14 a)}{7 (15+k_\Gamma-2a)}$, 
 & $\frac{7-2A}{8-A}$\\
$\nu_m<\nu<\nu_a$   &  ${5 \over 2}$  &   $\frac{-2a+10}{8-a}$ & $\frac{-2A+10}{8-A}$ & $ \frac{135-28a -5 k_\Gamma}{7(15+k_\Gamma-2a)}$ & $\frac{5-2A}{4-A}$\\
$\nu_a<\nu<\nu_c$   &  ${-{p-1 \over 2}}$   &  $\frac{-6p-3a+6}{8-a}$
  & $\frac{-6p-3A+6}{8-A}$ & $-\frac{3(7-7k_\Gamma+14a+27p-13k_\Gamma p)}{7(15+k_\Gamma-2a)}$ & $-\frac{3(A+p)}{8-A}$\\
$\nu>\nu_c$   &  ${-{p\over 2}}$   &   $\frac{-6p-2a+4}{8-a}$  & $\frac{-6p-2A+4}{8-A}$ & ${\rm cutoff}$ & {\rm cutoff} \\

\hline
ISM & fast cooling \\
\hline
$\nu<\nu_a$   &  ${2}$  &   $1$ &  $1$  &  ${\rm N/A}$  & ${\rm N/A}$ \\
$\nu_a < \nu<\nu_c$   &  ${1\over 3}$  &   $\frac{4-11a}{3(8-a)}$ &  $\frac{4-11A}{3(8-A)}$  &  ${\rm N/A}$  & ${\rm N/A}$ \\
$\nu_c<\nu<\nu_m$   &  $-{1\over 2}$  &  $\frac{-2(a+1)}{8-a}$  & $\frac{-2(A+1)}{8-A}$ & {\rm cutoff}   & {\rm cutoff}  \\
$\nu>\nu_m$   &  $-{p\over 2}$   &   $\frac{-6p-2a+4}{8-a}$      &  $\frac{-6p-2A+4}{8-A}$  &{\rm cutoff}   &{\rm cutoff} \\

\hline
Wind & slow cooling \\
\hline
$\nu < \nu_a$ & ${2}$   &   $\frac{4-2a}{4-a}$ &  $\frac{4-2A}{4-A}$ & $\frac{13-14a+22k_\Gamma}{7(3+k_\Gamma-a)}$ & $\frac{5-2A}{4-A}$\\
$\nu_a< \nu<\nu_m$   &  ${1\over 3}$   &   $-\frac{a}{3(4-a)}$ &  $ -\frac{A}{3(4-A)}$ & $-\frac{7a+5k_\Gamma+30}{21(3+k_\Gamma-a)}$ & $-\frac{5+A}{3(4-A)}$\\
$\nu_m<\nu<\nu_c$   &  $-{p-1\over 2}$  &   $-\frac{(6-a)p+a-2}{2(4-a)}$    &   $-\frac{(6-A)p+A-2}{2(4-A)}$ & $-\frac{7(1+k_\Gamma+a)+(39-11k_\Gamma-7a)p}{14(3+k_\Gamma-a)}$ & $-\frac{3}{4-A}-\frac{p-1}{2}$\\
$\nu>\nu_c$   &  $-{p\over 2}$  &   $-\frac{(6-a)p+2a-4}{2(4-a)}$   &  $-\frac{(6-A)p+2A-4}{2(4-A)}$   &   {\rm cutoff} &  {\rm cutoff}\\
\hline
$\nu < \nu_m$ & ${2}$   &   $\frac{4-2a}{4-a}$ &  $\frac{4-2A}{4-A}$ & $\frac{13-14a+22k_\Gamma}{7(3+k_\Gamma-a)}$ & $\frac{5-2A}{4-A}$\\
$\nu_m < \nu<\nu_a$   &  ${5\over 2}$   &   $\frac{14-5a}{2(4-a)}$ &  $ \frac{14-5A}{2(4-A)}$ & $\frac{65-35a+33k_\Gamma}{14(3-a+k_\Gamma)}$ & $\frac{14-5A}{2(4-A)}$\\
$\nu_a<\nu<\nu_c$   &  $-{p-1\over 2}$  &   $-\frac{(6-a)p+a-2}{2(4-a)}$    &   $-\frac{(6-A)p+A-2}{2(4-A)}$ & $-\frac{7(1+k_\Gamma+a)+(39-11k_\Gamma-7a)p}{14(3+k_\Gamma-a)}$ & $-\frac{3}{4-A}-\frac{p-1}{2}$\\
$\nu>\nu_c$   &  $-{p\over 2}$  &   $-\frac{(6-a)p+2a-4}{2(4-a)}$   &  $-\frac{(6-A)p+2A-4}{2(4-A)}$   &   {\rm cutoff} &  {\rm cutoff}\\
\hline
Wind & fast cooling  \\
\hline
$\nu<\nu_a$   &  $2$   &   $\frac{8-3a}{4-a}$ & $\frac{8-3A}{4-A}$ &  ${\rm N/A}$ &  ${\rm N/A}$   \\
$\nu_a<\nu<\nu_c$   &  ${1\over 3}$   &   $-\frac{8-a}{3(4-a)}$ & $-\frac{8-A}{3(4-A)}$ &  ${\rm N/A}$ &  ${\rm N/A}$   \\
$\nu_c<\nu<\nu_m$   &  $-{1\over 2}$  &   $-\frac{2+a}{2(4-a)}$  & $-\frac{2+A}{2(4-A)}$  &{\rm cutoff}  &{\rm cutoff} \\
$\nu>\nu_m$   &  $-{p\over 2}$  &   $-\frac{(6-a)p+2a-4}{2(4-a)}$    & $-\frac{(6-A)p+2A-4}{2(4-A)}$  & {\rm cutoff}   &{\rm cutoff} \\
\hline
\end{tabular}
\label{Tab:alpha-beta}
\begin{minipage}{2\columnwidth}
    \small Note: A thin shell approximation has been used for the reverse shock. The light curve slopes $\alpha$ are the ones after the deceleration time of the material at $\theta=\theta_{\rm j}$.
\end{minipage}
\label{tab:relation}
\end{table*}

\subsection{Reverse shock emission}\label{sec:RS}

For the matter-dominated structured jet wing, the reverse shock emission could be also important. Since the bulk Lorentz factor of the wing is smaller than that of the core, the deceleration epoch of the matter-dominated wing is usually at a later time. Observationally, the detection of the jet break in the TeV band suggests that the contrbution of the RS emission is initially not significant. As a result, the reverse shock emission from the matter-dominated wing is likely in the thin shell regime, i.e., $\Delta/c < t_{\rm dec,w}$, where $\Delta$ is the width of the  wing shell and $t_{\rm dec,w}$ is the deceleration time of the wing. In the following, we only consider the thin shell case. We also only consider the post-shock crossing phase. Before shock crossing, the wing materials are in the coasting phase, so that there is no change of emission contribution to the observer from different rings in the structured jet. The lightcurve behavior should be very similar to the uniform jet case\footnote{There should be a short transition phase when some rings of the structured jet is decelerated but some others are not, which would produce a new segment. We refrain from discussing these complications with the focus on interpreting the afterglow data of this GRB.}.

\subsubsection{Constant density medium}

For the reverse shock, the angular profile of the Lorentz factor of the shocked region (region 3) follows 
\begin{equation}
\Gamma_3 (\theta) =\Gamma_0 (\theta) \left[ \frac{r(\theta)}{r_{\rm dec}(\theta)}\right]^{-g},
\label{eq:Gamma3}
\end{equation}
where $g=2$ for a constant density medium in the thin-shell approximation \citep{kobayashisari00}.  Here $r_{\rm dec}(\theta)$ is the deceleration radius of the ejecta in the ring at $\theta$, which scales as 
\begin{equation}
r_{\rm dec}(\theta)\propto \epsilon(\theta)^{1/3}\Gamma_0(\theta)^{-2/3}.
\label{eq:r_dec}
\end{equation}
{Using $r(\theta)=2\Gamma_3(\theta)^2 c t$, Equation (\ref{eq:Gamma3}) can be written as
\begin{eqnarray}
 \Gamma_3(\theta) &\propto &
 \left[\Gamma_0(\theta) \right]^{\frac{1-2g/3}{1+2g}} [\epsilon(\theta)]^{\frac{g}{3(1+2g)}} t^{-\frac{g}{1+2g}} \\
 &\propto & [\Gamma_0(\theta)]^{-1/15} [\epsilon(\theta)]^{2/15} t^{-2/5}~~{\rm (for~}g=2).
\end{eqnarray}
Note that for reverse shock emission with $g=2$, the dynamics depends on the angular structures of both energy ($\epsilon(\theta) \propto \theta^{-k_\epsilon}$) and Lorentz factor ($\Gamma_0(\theta) \propto \theta^{-k_\Gamma}$). This is different from the forward shock case where the $\Gamma_0(\theta)$ profile does not enter the problem. Indeed, if we plug in $g=3/2$ (the scaling law of region 2), the $\Gamma_0(\theta)$ dependence disappears. }

As the dynamical evolution for $k_\Gamma\le 1$ and  $k_\Gamma> 1$ are different, we consider the two cases separately. 

\begin{center}
    {Case I: $k_\Gamma\le 1$}
\end{center}

In this case, one has $\Gamma_3(\theta)\propto \theta^{-1}$. 
Plugging in $\epsilon(\theta)$ and $\Gamma_0(\theta)$, one obtains
\begin{eqnarray}
\Gamma_3 & \propto & t^{-\frac{6}{15+k_\Gamma-2 k_\epsilon}}, \\
\theta & \propto & t^{\frac{6}{15+k_\Gamma-2 k_\epsilon}}, \\
\Gamma_0(\theta) & \propto & \theta^{-k_\Gamma}\propto  t^{-\frac{6k_\Gamma}{15+k_\Gamma-2k_\epsilon}}, \\
\epsilon(\theta) & \propto & \theta^{-k_\epsilon} \propto  t^{-\frac{6 k_\epsilon}{15+k_\Gamma-2k_\epsilon}}, \\
{\bar E}_{\rm iso}(\theta) & \propto & \theta^{-a} \propto  t^{-\frac{6 a}{15+k_\Gamma-2k_\epsilon}}.
\end{eqnarray}
These are general scaling laws to solve the problem. However, for $k_\epsilon \leq 2$, one has $a=k_\epsilon$, which can significantly simplify the derivations. Since for this burst $k_\epsilon < 2$ is always satisfied, in the following we replace $k_\epsilon$ by $a$ in all the derivations and do not differentiate $\epsilon(\theta)$ and $\bar E_{\rm iso}(\theta)$ and denote them with $E_{\rm iso}(\theta)$ universally. Please keep in mind that the scaling laws become much more complicated when $k_\epsilon > 2$.

The reverse shock emission peaks at the deceleration time $t_{\rm dec}(\theta)\propto r_{\rm dec}(\theta)/[2\Gamma_0^2(\theta)]\propto [E_{\rm iso}(\theta)]^{1/3} [\Gamma_0(\theta)]^{-8/3}$. The peak flux density at this time is $F_{\rm \nu,max}^{\rm rs}(t_{\rm dec})\propto E_{\rm iso} \Gamma_0$ and it decays with time roughly as $F_{\rm \nu,max}^{\rm rs}(t)=F_{\rm \nu,max}^{\rm rs}(t_{\rm dec}) (t/t_{\rm dec})^{-34/35}$ \citep{kobayashi00}.
We then have
\begin{eqnarray}
 F_{\rm \nu,max}^{\rm rs}(t) & \propto & [E_{\rm iso}(\theta)]^{\frac{139}{105}}[\Gamma_0(\theta)]^{-\frac{167}{105}} t^{-\frac{34}{35}} \nonumber \\
 & \propto & t^{-\frac{6(17-10k_\Gamma+7a)}{7(15+k_\Gamma-2a)}}. 
\end{eqnarray}
The reverse shock synchrotron frequency for the minimum Lorentz factor at the deceleration time is $\nu_m^{\rm rs}(t_{\rm dec}) \propto \Gamma_0^2 $  and it decays with time roughly as $\nu_{m}^{\rm rs}(t)\propto \nu_{m}^{\rm rs}(t_{\rm dec})(t/t_{\rm dec})^{-54/35}$. This gives
\begin{eqnarray}
    \nu_{m}^{\rm rs} & \propto & {E_{\rm iso}(\theta)}^{\frac{18}{35}} {\Gamma_0(\theta)}^{-\frac{74}{35}} t^{-\frac{54}{35}} \nonumber \\
    &\propto & t^{-\frac{6(27-13k_\Gamma)}{7(15+k_\Gamma-2a)}}.
\end{eqnarray}
The cooling break frequency   at the deceleration time  is $\nu_{c}^{\rm rs}\propto E_{\rm iso}^{-2/3} \Gamma_0^{4/3}$. After shock crossing, since no new electrons are accelerated, this frequency evolves as a cutoff frequency and decays following the same scaling law as $\nu_m$, i.e. $\nu_{\rm cut}^{\rm rs} \propto (t/t_{\rm dec})^{-54/35}$. This finally gives \citep{Gaohe2013}
\begin{eqnarray}
    \nu_{\rm cut}^{\rm rs} & \propto & [E_{\rm iso}(\theta)]^{-\frac{16}{105}} [\Gamma_0(\theta)]^{-\frac{292}{105}} t^{-\frac{54}{35}} \nonumber \\
    & \propto & t^{-\frac{2(81-53 k_\Gamma + 14a)} {7(15+k_\Gamma-2a)}}.
\end{eqnarray}
The self-absorption frequency is \citep{Gaohe2013}
\begin{eqnarray}
    \nu_a^{\rm rs} & \propto &[E_{\rm iso}(\theta)]^{\frac{69}{175}} [\Gamma_0(\theta)]^{\frac{8}{175}} t^{-\frac{102}{175}} \nonumber \\
    & \propto & t^{-\frac{-3(14a+10 k_\Gamma + 102)}{35(15+k_\Gamma-2a)}}
\end{eqnarray}
for $\nu_{a}^{\rm rs}< \nu_{m}^{\rm rs}<\nu_{\rm cut}^{\rm rs}$, and
\begin{eqnarray}\label{Eq.nua1}
     \nu_{a}^{\rm rs} & \propto & [E_{\rm iso}(\theta)]^{\frac{2(9p+29)}{35(p+4)}}[\Gamma_0(\theta)]^{-\frac{74p+44}{35(p+4)}} t^{-\frac{54p+104}{35(p+4)}} \nonumber \\
     & \propto & t^{-\frac{28a-78pk_\Gamma-32k_\Gamma+162p+312}{7(p+4)(15+k_\Gamma-2a)}}.
\end{eqnarray}
for $\nu_{m}^{\rm rs}< \nu_{a}^{\rm rs}<\nu_{c}^{\rm rs}$. 
Then the light curve of the reverse shock emission is given by
\begin{equation}
F_\nu \propto \left\{
 \begin{array}{ll}
   t^{\frac{-2(14 a -22 k_\Gamma-27)}{7 (15+k_\Gamma-2a)}}, &\nu < \nu_a^{\rm rs} \\
   t^{-\frac{2(24-17k_\Gamma+21a)}{7(15+k_\Gamma-2a)}}, &\nu_a^{\rm rs} <\nu < \nu_m^{\rm rs} \\
   t^{-\frac{3(7-7k_\Gamma+14a+27p-13k_\Gamma p)}{7(15+k_\Gamma-2a)}}, & \nu_m^{\rm rs} < \nu <\nu_{\rm cut}^{\rm rs} \\
   {\rm cutoff}. & \nu >\nu_{\rm cut}^{\rm rs}
 \end{array}
    \right.
\end{equation}
or
\begin{equation}
F_\nu \propto \left\{
 \begin{array}{ll}
   t^{\frac{2(27+22 k_\Gamma-14 a)}{7 (15+k_\Gamma-2a)}}, &\nu < \nu_m^{\rm rs} \\
   t^{\frac{135-28a -5 k_\Gamma}{7(15+k_\Gamma-2a)}}, &\nu_m^{\rm rs} <\nu < \nu_a^{\rm rs} \\
   t^{-\frac{3(7-7k_\Gamma+14a+27p-13k_\Gamma p)}{7(15+k_\Gamma-2a)}}, & \nu_a^{\rm rs} < \nu <\nu_{\rm cut}^{\rm rs} \\
   {\rm cutoff}. & \nu >\nu_{\rm cut}^{\rm rs}
 \end{array}
    \right.
\end{equation}
in the slow-cooling case ($\nu_m^{\rm rs}<\nu_{\rm cut}^{\rm rs}$). At shock crossing time, it is very unlikely that $\nu_c^{\rm rs}$ is smaller than than $\nu_m^{\rm rs}$. Since after shock crossing, $\nu_m^{\rm rs}$ and $\nu_{\rm cut}^{\rm rs}$ follow the same $t$-dependence, the fast cooling case would not be relevant and we will not discuss it further.  Note that when $a=0$ and $k_\Gamma=0$, the above equations reduce to the reverse shock emission for  the uniform fireball case \citep {kobayashi00,Gaohe2013}.

For $a=1$ and $k_\Gamma=0$, the  flux decays  as  $t^{-({63+81p})/{91}}$ ($\sim t^{-2.7}$ for $p=2.3$) in the spectral regime $\nu_m^{\rm rs} < \nu <\nu_{\rm cut}^{\rm rs}$. This is steeper than that of the reverse shock emission from a uniform fireball, which predicts $t^{-2}$ \citep {kobayashi00}. For $a=1$ and $k_\Gamma=1$, the  flux decays  as  $t^{-3(p+1)/7}$ ($\sim t^{-1.4}$ for $p=2.3$) in the spectral regime $\nu_m^{\rm rs} < \nu <\nu_{\rm cut}^{\rm rs}$, which is flatter than that of the reverse shock emission from a uniform fireball. This is because large angle material contributes significantly to the reverse shock emission for a steep angular distribution of the Lorentz factor (i.e., a large $k_\Gamma$).

\begin{center}
    {Case II: $k_\Gamma> 1$}
\end{center}
In this case, one has $\Gamma_3(\theta)\propto \theta^{-k_\Gamma}$. Plugging in $\epsilon(\theta)$ and $\Gamma_0(\theta)$, one obtains
\begin{eqnarray}
\Gamma_3 & \propto & t^{-\frac{3}{8-a/k_\Gamma}}\propto t^{-\frac{3}{8-A}}, \\
\theta & \propto & t^{\frac{3}{(8k_\Gamma-a)}}, \\
\Gamma_0(\theta) & \propto & \theta^{-k_\Gamma}\propto  t^{-\frac{3}{8-A}}, \\
E_{\rm iso}(\theta) & \propto & \theta^{-a} \propto t^{-\frac{3A}{8-A}}.
\end{eqnarray}
One can then derive
\begin{eqnarray}
    F_{\rm \nu,max}^{\rm rs}(t) & \propto & [E_{\rm iso}(\theta)]^{\frac{139}{105}}[\Gamma_0(\theta)]^{-\frac{167}{105}} t^{-\frac{34}{35}} t^{-\frac{3A+3}{8-A}}. \\
    \nu_m^{\rm rs} &\propto&  [E_{\rm iso}(\theta)]^{\frac{18}{35}} [\Gamma_0(\theta)]^{-\frac{74}{35}} t^{-\frac{54}{35}}\propto t^{-\frac{6}{8-A}}.\\
    \nu_{\rm cut}^{\rm rs} &\propto& [E_{\rm iso}(\theta)]^{-\frac{16}{105}} [\Gamma_0(\theta)]^{-\frac{292}{105}} t^{-\frac{54}{35}}\propto t^{\frac{2A-4}{8-A}}.
\end{eqnarray}

The self-absorption break frequency is given by
\begin{eqnarray}
    \nu_a^{\rm rs} & \propto & [E_{\rm iso}(\theta)]^{\frac{69}{175}} [\Gamma_0(\theta)]^{\frac{8}{175}} t^{-\frac{102}{175}} \propto t^{-\frac{3(8+A)}{5(8-A)}} ,
\end{eqnarray}
for the case of $\nu_a^{\rm rs}<\nu_m^{\rm rs}< \nu_{\rm cut}^{\rm rs}$ and
\begin{eqnarray}\label{Eq.nua2}
    \nu_a^{\rm rs} & \propto & {E_{\rm iso}(\theta)}^{\frac{2(9p+29)}{35(p+4)}}{\Gamma_0(\theta)}^{-\frac{74p+44}{35(p+4)}} t^{-\frac{54p+104}{35(p+4)}} \nonumber \\
     & \propto & t^{-\frac{2(A+3p+10)}{(p+4)(8-A)}}
\end{eqnarray}
for the case of $\nu_m^{\rm rs}< \nu_a^{\rm rs}<\nu_{\rm cut}^{\rm rs}$.

The light curve of the reverse shock emission is given by
\begin{equation}
F_\nu \propto \left\{
 \begin{array}{ll}
   t^{\frac{7-2A}{8-A}}, & \nu < \nu_a^{\rm rs} \\
   t^{-\frac{3A+1}{8-A}}, &\nu_a^{\rm rs}< \nu < \nu_m^{\rm rs} \\
   t^{-\frac{3(A+p)}{8-A}}, & \nu_m^{\rm rs} < \nu <\nu_{\rm cut}^{\rm rs} \\
   {\rm cutoff}. & \nu >\nu_{\rm cut}^{\rm rs} 
 \end{array}
    \right.
\end{equation}
or
\begin{equation}
F_\nu \propto \left\{
 \begin{array}{ll}
   t^{\frac{7-2A}{8-A}}, & \nu < \nu_m^{\rm rs} \\
   t^{\frac{2(5-A)}{8-A}}, &\nu_m^{\rm rs}< \nu < \nu_a^{\rm rs} \\
   t^{-\frac{3(A+p)}{8-A}}, & \nu_a^{\rm rs} < \nu <\nu_{\rm cut}^{\rm rs} \\
   {\rm cutoff}. & \nu >\nu_{\rm cut}^{\rm rs} 
 \end{array}
    \right.
\end{equation}
For $A=1$, the flux decays  as $t^{-3(p+1)/7}$ ($\sim t^{-1.4}$ for $p=2.3$)  in the spectral regime $\nu_m^{\rm rs} < \nu <\nu_{\rm cut}^{\rm rs}$, which is shallower than that of the reverse shock emission from a uniform jet.

\subsubsection{Wind medium}
For the wind medium, the angular profile of the Lorentz factor of the shocked shell follows 
\begin{equation}
\Gamma_3 (\theta) =\Gamma_0 (\theta) \left[\frac{r(\theta)}{r_{\rm dec}(\theta)}\right]^{-g}
\end{equation}
where $g=1$ \citep{Gaohe2013} for a  reverse shock in the thin-shell approximation and $r_{\rm dec}$ is the deceleration radius of the ejecta in the wind medium, which scales as $r_{\rm dec}(\theta)\propto \epsilon(\theta)\Gamma_0(\theta)^{-2}$. 
Using $r(\theta)=2\Gamma_3(\theta)^2 c t$, we obtain
\begin{eqnarray}
 \Gamma_3(\theta) &\propto &
 \left[\Gamma_0(\theta) \right]^{\frac{1-2g}{1+2g}} [\epsilon(\theta)]^{\frac{g}{1+2g}} t^{-\frac{g}{1+2g}} \nonumber \\
 &\propto & [\Gamma_0(\theta)]^{-1/3} [\epsilon(\theta)]^{1/3} t^{-1/3}~~{\rm (for~}g=1).
\end{eqnarray} 
Again, for forward shock dynamics with $g=1/2$, the $\Gamma_0$-dependence disappears.

\begin{center}
    {Case I: $k_\Gamma\le 1$}
\end{center}
In this case, one has $\Gamma_3(\theta)\propto \theta^{-1}$, which gives
\begin{eqnarray}
 \Gamma_3 &\propto& t^{-\frac{1}{3+k_\Gamma-a}}, \\
 \theta &\propto& t^{\frac{1}{3+k_\Gamma-a}},\\
 \Gamma_0(\theta) &\propto& \theta^{-k_\Gamma}\propto  t^{-\frac{k_\Gamma}{3+k_\Gamma-a}},\\
 E_{\rm iso}(\theta) &\propto& \theta^{-a} \propto t^{-\frac{a}{3+k_\Gamma-a}}.
\end{eqnarray}

One then has
\begin{eqnarray}
    F_{\rm \nu,max} &\propto& {E_{\rm iso}(\theta)}^{\frac{23}{21}}{\Gamma_0(\theta)}^{-\frac{29}{21}} t^{-\frac{23}{21}}\propto t^{-\frac{23-2k_\Gamma}{7(3+k_\Gamma-a)}}.   \\
    \nu_m^{\rm rs} &\propto&  {E_{\rm iso}(\theta)}^{\frac{6}{7}}{\Gamma_0(\theta)}^{-\frac{24}{7}} t^{-\frac{13}{7}}\propto t^{-\frac{39-11k_\Gamma-7a}{7(3+k_\Gamma-a)}}.\\
    \nu_{\rm cut}^{\rm rs} &\propto& {E_{\rm iso}(\theta)}^{\frac{20}{7}}{\Gamma_0(\theta)}^{-\frac{66}{7}} t^{-\frac{13}{7}}\propto t^{-\frac{39-53k_\Gamma+7a}{7(3+k_\Gamma-a)}}.
\end{eqnarray}

The self-absorption break frequency is given by 
\begin{eqnarray}\label{Eq.nua3} 
    \nu_a^{\rm rs} &\propto& {E_{\rm iso}(\theta)}^{\frac{6p-4}{7(p+4)}}{\Gamma_0(\theta)}^{\frac{16-24p}{7(p+4)}} t^{-\frac{13p+24}{7(p+4)}} \nonumber\\
    & \propto& t^{-\frac{(39-11 k_\Gamma-7a)p+40k_\Gamma-28a+72}{7(p+4)(3+k_\Gamma-a)}} 
\end{eqnarray}
for the case of $\nu_m^{\rm rs}< \nu_a^{\rm rs}<\nu_{\rm cut}^{\rm rs}$, and 
\begin{eqnarray}
    \nu_a^{\rm rs} &\propto& {E_{\rm iso}(\theta)}^{-\frac{12}{35}}{\Gamma_0(\theta)}^{\frac{48}{35}} t^{-\frac{23}{35}} \propto t^{-\frac{(69+71 k_\Gamma-35a)}{35(3+k_\Gamma-a)}} .
\end{eqnarray}
for the case of $\nu_a^{\rm rs}< \nu_m^{\rm rs} <\nu_{\rm cut}^{\rm rs}$. Then the light curve of the reverse shock emission is given by
\begin{equation}
F_\nu \propto \left\{
 \begin{array}{ll}
   t^{\frac{13-14a+22k_\Gamma}{7(3+k_\Gamma-a)}}, & \nu < \nu_a^{\rm rs} \\
   t^{-\frac{7a+5k_\Gamma+30}{21(3+k_\Gamma-a)}}, &\nu_a^{\rm rs}< \nu < \nu_m^{\rm rs} \\
   t^{-\frac{7(1+k_\Gamma+a)+(39-11k_\Gamma-7a)p}{14(3+k_\Gamma-a)}}, & \nu_m^{\rm rs} < \nu <\nu_{\rm cut}^{\rm rs} \\
   {\rm cutoff}. & \nu >\nu_{\rm cut}^{\rm rs}
 \end{array}
    \right.
\end{equation}
or
\begin{equation}
F_\nu \propto \left\{
 \begin{array}{ll}
   t^{\frac{13-14a+22k_\Gamma}{7(3+k_\Gamma-a)}}, & \nu < \nu_m^{\rm rs} \\
   t^{\frac{65 + 33 k_\Gamma -35a}{14(3+k_\Gamma-a)}}, &\nu_m^{\rm rs} < \nu < \nu_a^{\rm rs} \\
   t^{-\frac{7(1+k_\Gamma+a)+(39-11k_\Gamma-7a)p}{14(3+k_\Gamma-a)}}, & \nu_m^{\rm rs} < \nu <\nu_{\rm cut}^{\rm rs} \\
   {\rm cutoff}. & \nu >\nu_{\rm cut}^{\rm rs}
 \end{array}
    \right.
\end{equation}
For $a=1$ and $k_\Gamma=0$, the  flux decays  as  $t^{-({7+16p})/{14}}$  ($\sim t^{-3.1}$ for $p=2.3$) in the spectral regime $\nu_m^{\rm rs} < \nu <\nu_{\rm cut}^{\rm rs}$. This is steeper than that of the reverse shock emission from a uniform fireball. For $a=1$ and $k_\Gamma=1$, the  flux decays  as  $t^{-(p+1)/2}$ ($\sim t^{-1.65}$ for $p=2.3$) in the spectral regime $\nu_m^{\rm rs} < \nu <\nu_{\rm cut}^{\rm rs}$, which is flatter than that of the reverse shock emission from a uniform fireball. 

\begin{center}
    {Case II: $k_\Gamma> 1$}
\end{center}
In this case, one has $\Gamma_3(\theta)\propto \theta^{-k_\Gamma}$, which gives
\begin{equation}
\Gamma_3\propto t^{-\frac{k_\Gamma}{4k_\Gamma-a}}\propto t^{-\frac{1}{4-A}},
\end{equation}
with $A\equiv a/k_\Gamma$.
\begin{eqnarray}
    \theta &\propto& t^{\frac{1}{4-A}},\\
    \Gamma_0(\theta) &\propto& \theta^{-k_\Gamma}\propto  t^{-\frac{1}{4-A}},\\
    E_{\rm iso}(\theta) &\propto& \theta^{-a} \propto t^{-\frac{A}{4-A}}.
\end{eqnarray}
One then has
\begin{eqnarray}
    F_{\rm \nu,max} &\propto& {E_{\rm iso}(\theta)}^{\frac{23}{21}}{\Gamma_0(\theta)}^{-\frac{29}{21}} t^{-\frac{23}{21}}\propto t^{-\frac{3}{4-A}}.\\
    \nu_m^{\rm rs} &\propto&  {E_{\rm iso}(\theta)}^{\frac{6}{7}}{\Gamma_0(\theta)}^{-\frac{24}{7}} t^{-\frac{13}{7}}\propto t^{-1}.\\
    \nu_{\rm cut}^{\rm rs} &\propto& {E_{\rm iso}(\theta)}^{\frac{20}{7}}{\Gamma_0(\theta)}^{-\frac{66}{7}}  t^{-\frac{13}{7}}\propto t^{\frac{2-A}{4-A}}.
\end{eqnarray}

The self-absorption break frequency is given by 
\begin{equation}\label{Eq.nua4} 
    \nu_a^{\rm rs}\propto {E_{\rm iso}(\theta)}^{\frac{6p-4}{7(p+4)}}{\Gamma_0(\theta)}^{-\frac{16-24p}{7(p+4)}}  t^{-\frac{13p+24}{7(p+4)}}\propto t^{-1} ,
\end{equation}
for the case of $\nu_m^{\rm rs}< \nu_a^{\rm rs}<\nu_c^{\rm rs}$, and 
\begin{equation}
    \nu_a^{\rm rs}\propto {E_{\rm iso}(\theta)}^{-\frac{12}{35}}{\Gamma_0(\theta)}^{\frac{48}{35}}  t^{-\frac{23}{35}}\propto t^{-1} 
\end{equation}
for the case of $\nu_a^{\rm rs}< \nu_m^{\rm rs} <\nu_c^{\rm rs}$. The light curve of the reverse shock emission is then given by
\begin{equation}
F_\nu \propto \left\{
 \begin{array}{ll}
   t^{\frac{5-2A}{4-A}}, &\nu < \nu_a^{\rm rs} \\
   t^{-\frac{5+A}{3(4-A)}}, &\nu_a^{\rm rs}< \nu < \nu_m^{\rm rs} \\
   t^{-\frac{3}{4-A}-\frac{p-1}{2}}, &  \nu_m^{\rm rs}   < \nu <\nu_{\rm cut}^{\rm rs} \\
   {\rm cutoff}. & \nu >\nu_{\rm cut}^{\rm rs}
 \end{array}
    \right.
\end{equation}
or
\begin{equation}
F_\nu \propto \left\{
 \begin{array}{ll}
   t^{\frac{5-2A}{4-A}}, & \nu < \nu_m^{\rm rs} \\
   t^{\frac{11-5A}{2(4-A)}}, &\nu_m^{\rm rs} < \nu < \nu_a^{\rm rs} \\
   t^{-\frac{3}{4-A}-\frac{p-1}{2}}, &\nu_a^{\rm rs} < \nu < \nu_{\rm cut}^{\rm rs} \\
   {\rm cutoff}. & \nu >\nu_{\rm cut}^{\rm rs}
 \end{array}
    \right.
\end{equation}
for slow cooling. We also do not consider the case of fast cooling because it is usually not relevant\footnote{The condition $\nu_c^{\rm rs} < \nu_m^{\rm rs}$ may be satisfied at the shock crossing radius for a small parameter regime, which we do not deal with in this paper. }.
For $A=1$, the flux decays  as $t^{-(p+1)/2}$   in the spectral regime $\nu_m^{\rm rs} < \nu <\nu_{\rm cut}^{\rm rs}$, which is shallower than that of the reverse shock emission from a uniform jet.

\section{The two-component jet model}\label{sec:2-comp}

In this section, we discuss in detail the motivations of introducing the two-component jet delineated in Section \ref{sec:framework}. 

\subsection{Poynting-flux-dominated narrow pencil-beam jet}\label{sec:narrow}

The arguments in favor of a Poynting-flux-dominated narrow jet include the following:

\subsubsection{Non-detection of thermal component in the MeV spectrum during bright pulses}\label{sec:thermal}

The spectra of prompt emission during bright pulses are consistent with having a synchrotron radiation origin only \citep{BOAT-yang}. Since the fireball model usually predicts a bright thermal component in association with the non-thermal component \citep{meszarosrees00,peer06}, the non-detection of thermal emission can be used to place a lower limit on the magnetization parameter $\sigma$ of the outflow \citep{zhangpeer09}. For GRB 221009A, \cite{BOAT-yang} has applied a parametrization method to derive $\sigma > 45$ at the photosphere. In the following, we apply a more rigorous derivation to make the point.

Let us assume a relativistic GRB ``wind'' with an (isotropic) luminosity  $L_w = 10^{54} \ {\rm erg \ s^{-1}} L_{w,54}$, which is comparable to the observed isotropic $\gamma$-ray peak luminosity. If the outflow is a matter-dominated fireball, the critical specific enthalpy is \citep{meszarosrees00}
    \begin{eqnarray}
        \eta_* & \equiv & \left(\frac{L_w \sigma_T {\cal Y}}{8\pi m_p c^3 R_0} \right)^{1/4}
         \simeq  870 (L_{w,54} {\cal Y})^{1/4} R_{0,9}^{-1/4}, 
    \label{eq:eta*}
    \end{eqnarray}
    where ${\cal Y} \gtrsim 1$ is the lepton multiplicity, and $R_0 = (10^{9} \ {\rm cm}) R_{0,9}$ is the size of the central engine from a ``re-born'' fireball produced when the jet breaks out from the progenitor star. Since the maximum Lorentz factor achievable in a fireball is $\eta_*$, one always has $\Gamma_0 \leq \eta_*$. Comparing Eq.(\ref{eq:Gamma0}) and Eq. (\ref{eq:eta*}), one finds that the condition for $\Gamma_0 < \eta_*$ gives $R_0 < (2.5\times 10^9 \ {\rm cm}) L_{w,54} {\cal Y} (\Gamma_0/600)^{-4}$. The temperature at the central engine reads $T_0 \simeq (L_w/4\pi R_0^2 g_0 \sigma_B)^{1/4} 
    \simeq (4.7\times 10^{9} \ {\rm K \ or \ 405 \ {keV}}) L_{w,54}^{1/4} R_{0,9}^{-1/2}$. Assuming\footnote{Strictly speaking, a matter dominated fireball should have the afterglow-measured Lorentz factor smaller than the specific enthalpy at the central engine, i.e. $\Gamma_0 < \eta$ because of the radiation energy loss during the prompt emission phase (see \cite{zhang21} for a detailed treatment). } $\eta \sim \Gamma_0$, the expected photosphere component should have a temperature of \citep{meszarosrees00}
    \begin{eqnarray}
        T_{ph}= T_0 \left(\frac{\eta}{\eta_*}\right)^{8/3} 
        \simeq (1.7\times 10^9 \ {\rm K \ or \ 150 \ keV}), 
    \label{eq:Tph}
    \end{eqnarray}
    and a luminosity of \citep{meszarosrees00}
    \begin{eqnarray}
        L_{ph}=L_w \left(\frac{\eta}{\eta_*}\right)^{8/3}. 
        \label{eq:Lph}
    \end{eqnarray}
    Such a bright thermal component is difficult to hide from the observed spectrum. Taking the brightest epoch ($T_0+231-232 {\rm s}$) when the non-thermal emission is strongest (so that the thermal component may be over-shone), one can proceed with a most conservative constraint. At this epoch, following spectral parameters are derived \citep{BOAT-an}: the peak luminosity $L_{\rm peak} \sim 1.14 \times 10^{54} \ {\rm erg \ s^{-1}}$, peak energy $E_p \sim 1600$ keV, and low-energy photon index $\alpha \sim -0.93$. One can estimate the luminosity at 150 keV based on the $\nu F_\nu$ spectrum, which gives $L(150 \ {\rm keV}) \sim L_{\rm peak} (150/1600)^{2+\alpha} \sim 0.9 \times 10^{53} \ {\rm erg \ s^{-1}}$, which is a few time smaller than the  predicted value of photosphere luminosity $L_{ph}$ (Eq.(\ref{eq:Lph})). We conclude that the jet that is responsible for the prompt emission must be magnetically dominated even under the most conservative condition. 

\subsubsection{Non-detection of TeV emission during the prompt emission phase}
During the prompt emission phase, the  non-detection of TeV emission by LHAASO places an upper limit  on the flux ratio between the TeV and MeV emissions, $F_{\rm TeV}/F_{\rm MeV}\le 2\times10^{-5}$ \citep{BOAT-cao}. It is found that, the internal $\gamma\gamma$ absorption does not lead to an exponential attenuation of TeV photons in the framework of the internal shock scenario \citep{Granot2008,Aoi2010,Hascoet2012,Dai2023}. Instead, it is found that the superposition of spectra of different pulses with different cut-off energies results in a power-law-like spectrum for the time-integrated emission \citep{Aoi2010,Dai2023}. Therefore, the attenuation due to $\gamma\gamma$ absorption alone cannot explain this flux limit ratio of GRB 221009A, and  a low ratio of synchrotron self-Compton (SSC) to synchrotron emission outputs is required \citep{Dai2023}. Detailed calculations suggest that $\epsilon_{B}\gtrsim 10 \epsilon_{e}$ is needed to explain the the flux ratio between the TeV and MeV emissions \citep{Dai2023}.  This points to a Poynting-flux dominated outflow in which the radiation is powered by magnetic dissipation. 

\subsubsection{Non-detection of reverse shock emission from the bright core component}

A less stringent constraint is the non-detection of cross IC emission between the electrons and photons from the FS and RS in the $ \gtrsim 100$ GeV band (lower energy band of LHAASO) \citep{wang01,wang01b}. In particular, the IC of FS electrons off RS synchrotron photons is the most probable process to make these high-energy photons detectable by LHAASO. Assume that the narrow jet is matter dominated, one can estimate the flux ratio between this component and the FS SSC component, $R_{\rm F} \equiv F_\nu^{\rm IC-f,r} / F_\nu^{\rm SSC-f,f}$. This can be simply estimated using the following scaling relations between RS and FS emission at the RS shock crossing time. Assuming the same micro-physics parameters, one has \citep{kobayashizhang03a} $F_{\rm \nu,max}^{\rm rs}/F_{\rm \nu,max}^{\rm fs} \sim \Gamma_\times$ and $\nu_m^{\rm rs}/\nu_m^{\rm fs} \sim \Gamma_\times^{-2}$. Because $F_{\rm \nu,max}^{\rm IC} = \tau_e F_{\rm \nu.max}^{\rm syn}$ for both FS and RS seed photons and because $\tau_e$ is the same for the same FS electrons, one gets $F_{\rm \nu,max}^{\rm IC-f,r} / F_{\rm \nu,max}^{\rm SSC-f,f} \sim \Gamma_\times$ and $\nu_m^{\rm IC-f,r}/\nu_m^{\rm SSC-f,f} \sim \Gamma_\times^{-2}$. Assume that the lower LHAASO band is in the $\nu_m^{\rm IC-f,r} < \nu < \nu_c^{\rm IC-f,r}$ and $\nu_m^{\rm SSC-f,f} < \nu < \nu_c^{\rm SSC-f,f}$ band, one finally obtains
\begin{equation}
R_{\rm F} = \frac{F_{\rm \nu,max}^{\rm IC-f,r}} {F_{\rm \nu,max}^{\rm SSC-f,f}} \left(\frac{\nu_m^{\rm IC-f,r}}{\nu_m^{\rm SSC-f,f}} \right)^{\frac{p-1}{2}} \sim \Gamma_\times^{2-p}.
\end{equation}
Take $\Gamma_\times \sim 600$ and $p\sim 2.3$ as inferred from the data. This ratio is $R_{\rm F} \sim 0.147$, suggesting that the IC of RS photons by FS emission would give an extra contribution of about 1/7 of the FS SSC at the LHAASO afterglow peak time. It is unclear whether such a contribution can be ruled out from the data, but the data is certainly consistent with the non-detection of such a contribution.

\subsubsection{Non-detection of neutrino emission from the burst}

IceCube used a variety of methods to search for neutrinos in coincidence with  GRB 221009A over several time windows during the precursor, prompt and afterglow phases of the GRB \citep{BOAT-IceCube}. The analysis covers the energy range from MeV to PeV. The non-detection places stringent upper limits on the neutrino emission from this source \citep{BOAT-ai,BOAT-murase,BOAT-liu}. In particular, IceCube  found no track-like events from the direction of the GRB in a time range of $-1\, {\rm hr}$ to $+2 \,{\rm hr}$
from the trigger time, placing a muon-neutrino upper limit of $E_{\nu}^2dN_{\nu}/dE_{\nu}=3.9\times 10^{-2}{\rm GeV cm^{-2}} $ at the 90\% confidence level, under the assumption of an $E_{\nu}^{-2}$  power-law
spectrum between 0.8 TeV and 1 PeV \citep{BOAT-IceCube}. Using this limit, \citet{BOAT-ai} found that the dissipative photosphere model is highly disfavored.  For the internal shock model, the ratio of the total dissipated energy that goes into the protons and electrons is constrained to be $\epsilon_p/\epsilon_e\lesssim 10$ \citep{BOAT-ai}.  \cite{BOAT-murase} also concluded that the photosphere model is disfavored and the dissipation radius should be large. \citet{BOAT-liu} studied the electromagnetic cascade accompanying the neutrino production and found that the
constraint from the GeV observation  is stronger than that from the neutrino observation, with $\epsilon_p/\epsilon_e\lesssim 1$. All these constraints are consistent the ICMART model \citep{zhangyan11} that dissipates magnetic energy at a large radius from the central engine \citep{BOAT-ai,BOAT-murase,BOAT-liu}, a conclusion shared with studying non-detection of neutrinos from other GRBs as well \citep{zhangkumar13}. The most recent study by the IceCube team reached the similar constraints \citep{BOAT-IceCube2}. 

\subsubsection{Sharp edge requires a Poynting-flux-dominated flow}
When a jet with an extremely power penetrates through a collapsing massive star progenitor, matter in the envelope tends to mix with the jet materials to form a hot cocoon surrounding the jet \citep{zhangw03}. If the narrow jet is matter dominated, it would be surrounded by a structured jet wing with no significant energy contrast due to jet-cocoon mixing \citep{zhangw04}. The fact that the narrow jet has a sharp edge as revealed by the LHAASO data \citep{BOAT-cao} strongly suggests that such mixing is not important. A mechanism to protect the narrow jet from the mixing effect is to have the jet confined by a strong, ordered magnetic field  \citep[e.g.][]{gottlieb20}. Since the jet is highly relativistic, a Poynting-flux-dominated jet can ensure that the magnetic pressure is stronger than the thermal pressure of the relativistically hot plasma inside and surrounding the jet.

\subsubsection{Narrow jet consistent with a Poynting-flux-dominated flow}

From numerical simulations \citep{2008TchekhovskoyMNRAS.388..551T} and analytical arguments \citep{chenzhang21}, a Poynting-flux-dominated jet tends to produce a narrow jet opening angle with $\theta_{\rm j} \Gamma \sim$ a few. Even though for typical GRB jets one has $\theta_{\rm j} \Gamma > 10$, the narrow jet of GRB 221009A has $\theta_{\rm j} \sim 0.01$ and $\Gamma \sim 600$ \citep{BOAT-cao}, which gives $\Gamma\theta_{\rm j} \sim 6$, consistent with the expectation of a Poynting-flux-dominated jet.

\subsection{Matter-dominated  structured jet wing}\label{sec:structured}
The arguments for the existence of a matter-dominated wide structured jet wing include the following:

\subsubsection{Forward shock emission from the structured jet explains late-time broad-band emission}

Whereas the temporal slopes in the rising and the early decay phases in the LHAASO band are consistent with a narrow jet being decelerated by a constant density medium \citep{BOAT-cao}, the late time broad-band decay slopes are shallower than the prediction of a post-jet-break phase but steeper than the prediction of an isotropic fireball. This is consistent with having a structured jet with decreasing energy per solid angle with angle \citep{BOAT-structured-jet,BOAT-kann,BOAT-Gill}.  

One can apply the analytical model developed in Section \ref{sec:structured-jet} to interpret the afterglow data. The detailed fitting to the data have been carried out by \cite{BOAT-structured-jet,BOAT-Gill} without the inclusion of the LHAASO data and by \cite{zheng23,ren23,BOAT-zhang2} with the inclusion of the LHAASO data. Here we simply present some analytical estimates. According to \cite{BOAT-structured-jet}, the X-ray spectral index ranges from $\beta \sim -0.65$ at 1 hour to $\sim -1.10$ at 32 days. The early phase corresponds to $\nu_m < \nu < \nu_c$ with $p \sim 2.3$. The observed temporal slope $\alpha \sim 1.52$ steepens to $\alpha \sim 1.66$ after $\sim 0.82$ days. 

Using the forward shock closure relation in a constant density medium, one gets the structure profile index $a \sim 0.96$ for the case of $k_\Gamma \leq 1$ or $A \sim 0.96$ for $k_\Gamma>1$ at small angles (corresponding to observing times before $\sim 0.82$  days), and $a \sim 1.18$ for $k_\Gamma \leq 1$ or $A \sim 1.18$ for $k_\Gamma>1$ at larger angles (corresponding to times after $\sim 0.82$  days).  These angular profile indices are consistent with that derived by \cite{BOAT-structured-jet}.

In the wind medium,  one gets the structure profile index $a \sim 0.2$ for the case of $k_\Gamma \leq 1$ or $A \sim 0.2$ for $k_\Gamma>1$ at small angles (corresponding to the observing time before $\sim 0.82$  days), and $a \sim 0.73$ for $k_\Gamma \leq 1$ or $A \sim 0.73$ for $k_\Gamma>1$ at larger angles (corresponding to the observing time after $\sim 0.82$  days). 
These angular profile indices are roughly consistent with that derived by \cite{BOAT-Gill}. 

\subsubsection{Reverse shock from structured jet wing explains early radio emission}

According to \cite{Bright2023}, the rapid follow-up radio observations of  GRB 221009A  reveals an optically thick rising component. They identified a spectral break likely due to synchrotron self-absorption, with the break frequency evolving as $\nu_{a} \propto t^{-1.08\pm 0.04}$ and the flux density at the break frequency evolving as $F_{\nu_a} \propto t^{-0.70 \pm 0.02}$  between 0.283 d
and 1.7 d after the burst, which is not expected from the standard afterglow models, even for reverse shock emission. \cite{BOAT-Gill} numerically fit the broad-band afterglow data using a structured jet model with reverse shock contribution, but this particular feature was not interpreted.

Since the Poynting-flux-dominated core does not provide a reverse shock emission component \citep{zhangkobayashi05}, one should only focus on the reverse shock emission contribution from the matter dominated wing. This component provides a natural interpretation of the $\nu_a$ feature in the radio band and its evolution \citep{Bright2023,BOAT-laskar}. The radio spectrum is consistent with  the case of $\nu_m^{\rm rs}< \nu_a^{\rm rs}<\nu_{\rm cut}^{\rm rs}$.
In the case of a constant density medium, 
with $a=0.96$ and $k_\Gamma=0$, we get $\nu_a\propto t^{-1.23}$ (Eq.\ref{Eq.nua1}) and $F_{\nu_a} \propto t^{-1.9}$ (for $p=2.3$).  For $k_\Gamma>1$,  we get $\nu_a\propto t^{-\frac{2(A+3p+10)}{(p+4)(8-A)}} \propto t^{-0.81}$ (Eq.\ref{Eq.nua2}) and $F_{\nu_a} \propto t^{-0.87}$ with $A=0.96$. For a wind medium, on the other hand,  with $a \sim 0.2$ and $k_\Gamma=0$, we get  $\nu_a\propto t^{-1.24}$ (Eq.\ref{Eq.nua3}) and $F_{\nu_a} \propto t^{-1.62}$ (for $p=2.3$). For  $k_\Gamma>1$, we get $\nu_a\propto t^{-1}$ (Eq.\ref{Eq.nua4})and $F_{\nu_a} \propto t^{-0.79}$ with $A=0.2$ and the slopes are independent of the value of $p$. All these model predictions are consistent with the general trend of the observation, and the case of a wind medium with $A=0.2$ gives the best match with the observations. 

\subsubsection{Thermal emission from the quiet period}    

Even though there is no significant thermal component during bright prompt emission episodes \citep{BOAT-yang, BOAT-GBM}, a thermal component with $kT = 19.27 \pm 1.66$ keV was identified during the ``quiet period'' bridging the precursor and the bright emission episodes based on the Fermi GBM observations \cite{BOAT-GBM}. This thermal component may be interpreted as the off-beam photosphere emission of the matter-dominated structured jet wing. A self-consistent result can be reached if the line of sight is slightly off the jet axis. For example, if the line of sight is $0.25^{\rm o}$ from the axis, the closest approach to the matter-dominated wing is about $\theta_v=0.35^{\rm o}$. Suppose that the isotropic luminosity of the closest ring to the jet axis is $\sim 10^{53} \ {\rm erg \ s^{-1}}$ and that it moves with a Lorentz factor of $\sim 230$. The photosphere temperature along that direction is $T_{\rm ph}=T_0 (\Gamma/\eta_*)^{8/3} \sim 228 \,{\rm 
 keV} (230/490)^{8/3} \sim 30$ keV, where the relations of $T_0$ and $\eta_*$ in section \ref{sec:thermal} have been used. Along the line of sight, the temperature should be $T_{\rm obs} \simeq T_{\rm ph} ({\cal D}/\Gamma) \sim 20$ keV, where the Doppler factor ${\cal D}=[\Gamma(1-\beta\cos\theta_v)]^{-1} \simeq 155$. This is consistent with the observed temperature of the thermal component. This suggests that the matter dominated structured jet wing is already launched at least shortly after the precursor phase and continued for some time at least until the main dissipation episode started. 

\subsubsection{Spectral hardening in the TeV band}

Even though the SSC model can generally account for the global LHAASO data, the SSC spectrum predicts a softening at high-energy LHAASO band. The data instead shows a lack of such a turnover, which suggests the possible existence of another spectral component \citep{BOAT-cao2}. This hardening may be interpreted as the emergence of a hadronic component \citep[e.g.][]{wangk23}. One possibility is that it originates from the onset of proton synchrotron radiation in the reverse shock of the matter-dominated structured jet wing \citep{BOAT-zhang2}. 

\section{Implications}\label{sec:implications}
The observations of this unique burst raised several questions regarding the physical mechanism of this particular GRB and GRBs in general. 

\subsection{Jet launching and collimation}

One immediate question is how such a narrow Poynting-flux-dominated jet with a matter-dominated structured jet wing is launched and collimated. Studies of other long GRBs showed that most GRBs have wider jet opening angles \citep[e.g.][]{frail01,wangxg18}. This suggests that the progenitor of GRB 221009A may have unusual collimation than most others.  On the other hand, it may be possible that such narrow jets are not uncommon but are just missed for the majority of GRBs (see Section{\ref{sec:common}} below). {\bf Nonetheless}, for narrowly beamed Poynting-flux-dominated jets as observed in this burst, we speculate that the central engine, likely a hyper-accreting black hole that powers the jet via the Blandford-Znajek mechanism \citep{blandford77,lee00,lei13,lei17,liu17}, may have an unusually high $\sigma_0$ at the central engine. It could be possible that the progenitor star may be more extended or denser than other GRB progenitors so that an extra collimation may be achieved because of the pressure confinement \citep{lazzati12}.  The central BZ jet carries a strong magnetic barrier that keep baryons away, making a cleaner, narrower jet, whereas the surrounding matter-dominated structured jet is powered by neutrino-anti-neutrino annihilations that typically have a smaller power and heavier baryon loading \citep{lei13}. 

\subsection{Are such narrow-beam GRBs common?}\label{sec:common}

If such narrow-beam GRBs are not uncommon in GRB progenitors, another question is whether there is an observational signature for the very bright, narrow pencil beam jet viewed off-axis, and whether such a signature has been detected in the past from other GRBs. 

For a relativistic jet viewed at a larger angle, there is a ``closure relation'' among isotropic energy $E_{\rm iso}$, peak frequency or energy $E_p$, and the duration $T_{90}$ for the on-beam and off-beam observers. Such a closure relation has been studied within the context of fast radio bursts (FRBs) by \cite{zhang21b}. Here we study such closure relations relevant to GRBs. Unlike FRBs whose on-axis and off-axis observations are all made in the similar radio bands, GRB observations have the advantage of observing bursts in much wider energy bands and across much wider temporal domains. The closure relations are then much simpler. Let us assume that the on-axis Doppler factor is ${\cal D}^{\rm on} \simeq \Gamma$ (the bulk Lorentz factor of the narrow jet) and the off-axis Doppler factor is ${\cal D}^{\rm off} = [\Gamma(1-\beta\cos\theta_v)]^{-1}$ ($\theta_v$ is the angle between the line of the sight and the edge of the narrow jet). Considering that in the observer frame one has $E = {\cal D} E'$, $\nu={\cal D} \nu'$, and $t={\cal D}^{-1} t'$, where $E$, $\nu$, $t$ are energy, frequency, and time, respectively, and the primed quantities are measured in the comoving frame, one finally gets the simple closure relations
\begin{equation}
 \frac{E_{\rm iso}^{\rm on}}{E_{\rm iso}^{\rm off}} = \frac{E_p^{\rm on}}{E_p^{\rm off}} = \frac{T_{90}^{\rm off}}{T_{90}^{\rm on}} = \frac{{\cal D}^{\rm on}}{{\cal D}^{\rm off}}. 
\end{equation}
With this relation, one can speculate that some very bright early X-ray flares with fluences comparable to those of prompt emission (e.g. the one detected from GRB 050502B, \cite{falcone06}) could be regarded as off-axis narrow beam emission. This X-ray flare had a fluence of $\sim 10^{-6} \ {\rm erg \ cm^{-2}}$. There is no redshift measurement, but various indicators suggest that it likely has a relatively high redshift \citep[e.g.][]{afonso11}. Taking $z=5.2$ as suggested by \cite{afonso11}, the isotropic energy of the X-ray flare is $E_{\rm iso}^{\rm off} \sim 3\times 10^{53}$ erg, which is greater than the isotropic energy of prompt emission of that GRB ($\sim 4\times 10^{52}$ erg). The duration of the flare (starting from the trigger, with early rising phase buried beneath the X-ray afterglow) is $T_{90}^{\rm off} \sim 2000$ s \citep{falcone06}. The main emission of GRB 221009A has a duration of $T_{90}^{\rm on} \sim 60$ s (not including the pre-cursor and the late flare, which may not be observable for an off-axis observer) \citep{BOAT-an}. Considering its $E_{\rm iso}^{\rm on} \sim 10^{55}$ erg, the closure relation is roughly satisfied for $E_{\rm iso}$ and $T_{90}$ relations, with ${\cal D}^{\rm on} / {\cal D}^{\rm off} \sim 30$. The $E_p$ of the X-ray flare is not measured, but \cite{falcone06} detected an obvious spectral hardening during the giant X-ray flare phase with photon index harder than -2, suggesting that $E_p$ of the X-ray flare is above the XRT energy band, i.e. $>10$ keV. Noticing $E_p^{\rm on} \sim 1.2$ MeV \citep{BOAT-an}, the $E_p$ closure relation could be also well satisfied. We therefore suggest that GRB 050502B could be a GRB similar to GRB 221009A whose bright core is missed by the observer. Assuming that the narrow jet also has a bulk Lorentz factor of $\sim 600$, the viewing angle for GRB 050502B is $\theta_v \sim 0.73^{\rm o}$ outside the $0.6^{\rm o}$ cone, or $\sim 1.33^{\rm o}$ from the jet axis. This viewing angle is also very small. This suggests that for the majority of GRBs with a narrow pencil-beam jet, ${\cal D}^{\rm off}$ should be much smaller (drops below unity when $\theta_v > 3.3^{\rm o}$). This suggests that it is indeed possible that such narrow pencil-beam jets could present in a large sample of GRBs but have escaped detection. They may not be detected at late times in the afterglow phase, either, because the isotropic energy of line-of-sight emission is usually greater than the contribution from the narrow jet when it enters the line of sight. 

It is worth clarifying that we do not interpret all X-ray flares as off-beam narrow jets. The majority of X-ray flares should be still related to the late central engine activities, as supported by mounting observational evidence \citep{zhang06,liang06,uhm16a,jia16}, but the giant X-ray flares such as the one detected in GRB 050502B \citep{falcone06}, which requires significant energy budget \citep{wu13}, could be explained more efficiently by the off-beam narrow jet viewed at a small enough viewing angle. 

\section{Summary and Discussion}\label{sec:summary}

We have suggested that the BOAT GRB 221009B has two jet components: one narrow pencil-beam jet with an opening angle of $\sim 0.6^{\rm o}$ that is Poynting-flux-dominated, and a wider matter-dominated structured jet wing with a shallow energy-distribution profile. 

The arguments in favor of a Poynting-flux-dominated narrow jet include the following:
\begin{itemize}
    \item There is no thermal emission component detected during the bright MeV pulses; 
    \item There is no TeV detection during the prompt emission phase;
    \item There is no detction of the reverse shock component associated with the prompt emission jet;
    \item There is no detection of neutrinos from the burst;
    \item There is a clear jet break detected in the TeV lightcurve which suggests a sharp edge of the narrow jet;
    \item The narrow jet has a $\Gamma\theta_{\rm j}=$ a few, which is consistent with a Poynting flux dominated jet.
\end{itemize}
The arguments that call for a wider matter-dominated structured jet wing include the following:
\begin{itemize}
    \item The broad-band afterglow data starting from hours after the trigger could not be interpreted by the narrow jet seen from the TeV lightcurve, and requires an additional jet component. Its broad properties (including decay slope and closure relations) require a non-uniform angular structure. 
    \item The early radio emission cannot be accounted for by the forward shock model, but could be interpreted by the reverse shock emission of a structured jet. The existence of a reverse shock suggests that this structured jet wing is matter dominated.
    \item There is a thermal component during the quiet phase of prompt emission, which could be interpreted as the photosphere emission of the matter-dominated structured jet wing.
    \item There is a spectral hardening in TeV emission that could not be accounted for by the standard forward shock SSC model. One possible interpretation is to introduce proton synchrotron emission from the reverse shock of the matter-dominated structured jet.
\end{itemize}

To solve the problem, we have developed an analytical theoretical framework for structured jets. The model deals with both forward and reverse shock synchrotron emission of the structured jet during the deceleration phase. We emphasize that the effective isotropic energy within the $1/\Gamma$ cone rather than the isotropic energy in a particular ring in the structured jet may be used to more precisely study the problem. At large angles, the angular structure index $a = {\rm min} (k_\epsilon, 2)$ is consistent with the energy profile index $k_\epsilon$, but only when $k_\epsilon \leq 2$. At early epochs, the effective $E_{\rm iso}$ of the structured jet wing initially increases with $\theta$, which gives a more precise treatment of the problem. 

We did not derive the closure relations of SSC emission from the structured jet in this paper, but they could be properly derived following the similar procedure. In any case, similar to the synchrotron component, it is expected that the SSC lightcurves are steeper than that of a top hat jet. In our detailed numerical study \citep{zheng23}, we have included the SSC emission from the structured jet to fit the broadband data of this burst. Our results showed that the SSC component from the structured jet is not significant, which is consistent with the late-time TeV flux upper limit provided by H.E.S.S. observations \citep{BOAT-HESS}.

We also discussed the physical implications gained from this burst. The narrow jet may require a highly magnetized engine or a more extended stellar envelope to collimate the jet. On the other hand, we also argue that the narrow jets may not be rare among GRBs. We suggest that some bright X-ray flares (such as the one observed in GRB 050502B) with fluence comparable to or even greater than prompt emission could be emitted by a narrow jet viewed slightly outside the jet cone. For the majority of cases, the narrow beam could be completely missed by the observer. A narrow jet similar to the one observed in GRB 221009A may exist in many more GRBs but may have escaped detection. Within such a picture, the majority of GRBs are viewed within the structured jet cone. As inferred from the modeling of this burst, the jet structure of this second component is shallow, so that the general afterglow behavior may not differ too much from the top hat jet case, and the jet break usually measured around a day \citep{frail01,wangxg18} would correspond to the edge of this shallow structured jet. Nonetheless, the equal-arrival-time surface of such GRBs are different from the top hat case, which carries the imprints of the structured jet \citep[e.g.][]{beniamini20}. Imaging observations and polarization measurements of some nearby events may lead to a diagnosis of the jet structure, which would offer support or disfavor the universal structured jet scenario proposed in this paper.

\section*{Acknowledgements}
The authors thank an anonymous referee for helpful comments and the following colleagues for stimulative discussions on modeling this burst: Shunke Ai, Paz Beniamini, Eric Burns, Zhen Cao, Connery Chen, Liang Chen, Cui-Yuan Dai, Zi-Gao Dai, He Gao, Ramandeep Gill, Ali Kheirandish, Shigeo Kimura, Tanmoy Laskar, Ruo-Yu Liu, Peter M\'esz\'aros, Kohta Murase, Asaf Pe'er, Yuri Sato, Donggeun Tak, Z. Lucas Uhm, Kai Wang, Yu-Jia Wei, Bin-Bin Zhang, B. Theodore Zhang, and Yuan-Chuan Zou.







\end{document}